\documentclass[12pt]{article}
\usepackage{amsmath,amssymb,epsfig}
\makeatletter \@addtoreset{equation}{section} \makeatother
\renewcommand{\theequation}{\thesection.\arabic{equation}}
\newcommand{\ba}{\begin{array}}
\newcommand{\ea}{\end{array}}
\newcommand{\beq}{\begin{equation}}
\newcommand{\eeq}{\end{equation}}
\newcommand{\bea}{\begin{eqnarray}}
\newcommand{\eea}{\end{eqnarray}}

\def\bce{\begin{center}}
\def\ece{\end{center}}

\def\nonu{\nonumber}

\def\pa{\partial}

\def\be{\beta}

\def\eps6{{\displaystyle \mathop{\epsilon}^{6}}{}}

\def\nab6{{\displaystyle \mathop{\nabla}^{6}}{}}

\def\0{{\sst{(0)}}}
\def\1{{\sst{(1)}}}
\def\2{{\sst{(2)}}}
\def\3{{\sst{(3)}}}
\def\4{{\sst{(4)}}}
\def\5{{\sst{(5)}}}
\def\6{{\sst{(6)}}}
\def\7{{\sst{(7)}}}
\def\8{{\sst{(8)}}}

\def\ba{\begin{array}}
\def\ea{\end{array}}
\def\beq{\begin{equation}}
\def\eeq{\end{equation}}
\def\be{\begin{equation}}
\def\ee{\end{equation}}

\def\eps{\epsilon}

\def\ba{\begin{array}}
\def\ea{\end{array}}
\def\beq{\begin{equation}}
\def\eeq{\end{equation}}
\def\be{\begin{equation}}
\def\ee{\end{equation}}

\def\eps{\epsilon}

\newcommand{\bean}{\begin{eqnarray*}}
\newcommand{\eean}{\end{eqnarray*}}

\begin{document}
\thispagestyle{empty} \addtocounter{page}{-1}
\begin{flushright}
\end{flushright}

\vspace*{1.3cm}

\centerline{ \Large \bf  Meta-Stable Brane Configurations }
\vspace{.3cm} 
\centerline{ \Large \bf by Adding an Orientifold-Plane to Giveon-Kutasov } 
\vspace*{1.5cm}
\centerline{{\bf Changhyun Ahn} 
} 
\vspace*{1.0cm} 
\centerline{\it 
Department of Physics, Kyungpook National University, Taegu
702-701, Korea} 
\vspace*{0.8cm} 
\centerline{\tt ahn@knu.ac.kr} 
\vskip2cm

\centerline{\bf Abstract}
\vspace*{0.5cm}

In hep-th/0703135, they have found the type IIA
intersecting brane
configuration where there exist three NS5-branes, D4-branes and 
anti-D4-branes.
By analyzing the gravitational interaction for the D4-branes in the
background
of the NS5-branes, the phase structures in different regions of the
parameter space were studied in the context of classical string theory.
In this paper, by adding the orientifold 4-plane and 6-plane to the above brane
configuration,
we describe the intersecting brane configurations of type IIA 
string theory corresponding to the meta-stable nonsupersymmetric 
vacua of these gauge theories.  

\baselineskip=18pt
\newpage
\renewcommand{\theequation}
{\arabic{section}\mbox{.}\arabic{equation}}

\section{Introduction}

It is known that the NS5-brane configuration in type IIA string theory
where 
there exist two types of NS5-branes, i.e., NS5-brane(012345) and 
NS5'-brane(012389), preserves ${\cal N}=2$
supersymmetry in four dimensions \cite{GK98}.
By adding D4-branes(01236),  that are suspended between the
NS5-brane and the NS5'-brane, and 
anti-D4-branes($\overline{D4}$-branes), that are suspended between the
NS5-brane and the other NS5'-brane, into this system, the
supersymmetry is broken \cite{GK}. 
As the distance between the two NS5'-branes along 
one of the longitudinal directions of the NS5-brane 
becomes zero, this brane configuration with
D4- and $\overline{D4}$-branes can decay and the geometric 
misalignment between 
flavor D4-branes, which can be interpreted as a nontrivial F-term
condition in the gauge theory side, 
arises.  
Due to the presence of NS5-brane in this system, there 
is an attractive force between the tilted D4-branes and NS5-brane.
The explicit computation of DBI action for these D4-branes in the
background of NS5-brane is done by the work of \cite{GK} recently 
and this effect of the gravitational attraction 
leads to a curve for tilted D4-branes rather than a
straight line. The meta-stable vacua of \cite{ISS} appear in some
region of parameter space. 

In this paper, we focus on the new meta-stable brane configurations 
by adding an orientifold 4-plane and an orientifold 6-plane to 
the above brane configuration studied by \cite{GK}, along the line of 
\cite{FGU,OO,BGHSS,Ahn06}.
When the former is added, no extra NS5-branes or D-branes are needed.
However, when the latter is added, the extra NS5-branes or D-branes 
into the above
brane configuration are needed in order to have a product gauge group. 
All of these examples have very simple dual magnetic superpotentials 
which make it easier to analyze meta-stable brane 
configurations. 

In section 2, we review the type IIA brane configuration corresponding
to the electric theory based on the ${\cal N}=1$ $Sp(N_c) \times
SO(2N_c')$ 
gauge theory 
with a bifundamental and deform this theory by adding the mass term
for the bifundamental. 
Then we construct the dual magnetic theory which is 
${\cal N}=1$ $Sp(\widetilde{N}_c) \times SO(2N_c')$ gauge 
theory with corresponding dual
matter as well as gauge singlet for the first gauge group factor. 
We consider the nonsupersymmetric meta-stable
minimum  and present 
the corresponding intersecting brane configurations of type IIA string
theory.
We also discuss  the dual magnetic theory which is 
${\cal N}=1$ $Sp(N_c) \times SO(2\widetilde{N}_c')$ gauge 
theory briefly.

In section 3, we describe the type IIA brane configuration corresponding
to the electric theory based on the ${\cal N}=1$ $SU(N_c) \times
SU(N_c')$ 
gauge theory 
with  matters and deform this theory by adding the mass term
for the bifundamentals. 
Then we construct the  dual magnetic theory which is 
${\cal N}=1$ $SU(\widetilde{N}_c) 
\times SU(N_c')$ gauge 
theory with corresponding dual
matters as well as gauge singlet for the first gauge group factor. 
We consider the nonsupersymmetric meta-stable
minimum  and present 
the corresponding intersecting brane configurations of type IIA string
theory. We also consider the same gauge theory with different matters 
and  describe the nonsupersymmetric meta-stable brane configuration
from the  dual magnetic theory which is 
${\cal N}=1$ $SU(\widetilde{N}_c) \times SU(N_c')$ gauge 
theory.

In section 4,  we make some comments for the future directions.  

\section{When an O4-plane is added}

In this section, we add an orientifold 4-plane to the type IIA brane
configurations of \cite{GK} and construct new 
meta-stable brane configurations. 
  
\subsection{Electric theory}

The type IIA brane configuration corresponding to 
${\cal N}=1$ supersymmetric gauge theory with
gauge group
\bea
Sp(N_c) \times SO(2N_c')
\label{gaugegroup}
\eea
and a bifundamental $X$ that is in the representation 
$({\bf 2N_c, 2N_c'})$ under the gauge group (\ref{gaugegroup}) 
can be described by 
a middle NS5-brane(012345),
the left $NS5_L'$-brane(012389) and the right 
$NS5_R'$-brane(012389), $2N_c$-  and $2N_c'$-color D4-branes(01236) as
well as an 
$O4^{+}$-plane(01236) and an $O4^{-}$-plane(01236) we should add. 
We take the arbitrary number of color D4-branes with the constraint 
$N_c' \geq N_c+2$.
The $O4^{\pm}$-planes act as $(x^4,x^5,x^7,x^8,x^9) \rightarrow
(-x^4,-x^5,-x^7,
-x^8,-x^9)$ as usual 
and they have RR charge $\pm 1$ playing the role of $\pm 1$
D4-brane.
The bifundamental $X$  corresponds to 4-4 strings connecting 
the $2N_c$-color D4-branes with $2N_c'$-color D4-branes.

The middle NS5-brane is located at $x^6=0$ and we denote the $x^6$ 
coordinates for the $NS5_L'$-brane and $NS5_R'$-brane 
by $x^6=-y_1 (< 0)$ and
$x^6=y_2 (> 0)$
respectively, along the line of \cite{GK}. 
The $2N_c$ D4-branes and $O4^{+}$-plane 
are suspended between the middle
NS5-brane and $NS5_R'$-brane while 
the $2N_c'$ D4-branes and $O4^{-}$-plane 
are suspended between the $NS5_L'$-brane and the middle NS5-brane.
Moreover, there exist $O4^{+}$-plane(which will extend to $x^6=-\infty$) 
to the left side of
$NS5_L'$-brane
and $O4^{-}$-plane(which will extend to $x^6=+\infty$) to the right side of
$NS5_R'$-brane.
We draw this brane configuration in Figure 1A for the vanishing mass
for the bifundamental $X$ by inserting the appropriate
$O4^{\pm}$-planes
into the brane configuration of \cite{GK} 
\footnote{This is equivalent to the reduced brane
  configuration of \cite{Ahn07-2} if we remove D6-branes 
from \cite{Ahn07-2} completely.}. See also 
the relevant works appeared in 
\cite{Tatar,Ahn97,Hashiba,Ahn07-2}. The gauge group and matter content
of \cite{GK} are changed as above by orientifolding procedure to that theory. 

\begin{figure}[ht]
   \epsfxsize=4.0in 
\centerline{\epsffile{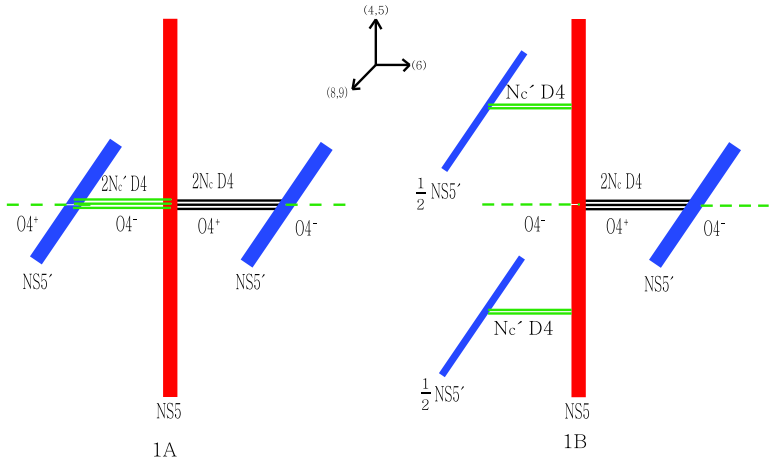}}
   \caption[FIG. \arabic{figure}.]{ 
The  ${\cal N}=1$ supersymmetric 
electric brane configuration for the gauge group $Sp(N_c) \times
SO(2N_c')$ and a bifundamental $X$  with vanishing(1A) and 
nonvanishing(1B) mass
for the bifundamental. The bifundamental $X$  corresponds to 4-4 
strings connecting 
the $2N_c$-color D4-branes with $2N_c'$-color D4-branes.}
\end{figure}

The gauge couplings of $Sp(N_c)$ and $ SO(2N_c')$
are given by a string coupling constant $g_s$, a string scale $\ell_s$ 
and the $x^6$ coordinates $y_i$ for two NS5'-branes through
\bea
g_{Sp}^2 = \frac{g_s \ell_s}{y_2}, \qquad 
g_{SO}^2 = \frac{g_s \ell_s}{y_1}
\nonu
\eea
respectively.
As $y_1$ goes to $\infty$ implying the change of 
the relative strength for the
two gauge couplings, the $SO(2N_c')$ gauge group becomes a
global symmetry and the theory leads to SQCD with the gauge group
$Sp(N_c)$ and $N_c'$ flavors(or $2N_c'$ fields) 
in the fundamental representation.
On the other hand, 
the opposite limit $y_2 \rightarrow \infty$ leads to SQCD with the
gauge group $SO(2N_c')$ with $2N_c$ fields in the fundamental
representation.

There is no superpotential in Figure 1A.
Let us deform this gauge theory. 
Displacing the two NS5'-branes relative each other in the $v \equiv
x^4 + i x^5$ direction corresponds to turning on a quadratic
mass-deformed superpotential
for the bifundamental $X$ as follows:
\bea
W = m X X \left( \equiv m \Phi \right)
\label{superelectric}
\eea
where a symplectic metric that has antisymmetric color 
indices \cite{Ahn07-2} 
is assumed in the $Sp(N_c)$ gauge group 
indices for $X X$, the $\Phi$
is a meson field
and the mass $m$ is given by geometrically
\bea
m = \frac{\Delta x}{2 \pi \alpha'} 
\left( = \frac{\Delta x}{\ell_s^2} \right).
\nonu
\eea
Half of $NS5_L'$-brane together with $N_c'$ color D4-branes 
is moving to the $+v$ direction and half of 
$NS5_L'$-brane  together with other $N_c'$ color D4-branes
is moving to $-v$ direction due to the O4-plane for
fixed $NS5_R'$-brane during this mass deformation. See also \cite{BFH}
for the splitting of branes on orientifold planes in the general context.
The splitting of $NS5_R'$-brane for fixed $NS5_L'$-brane can be
applied also and will be explained later in subsection 2.4. 
Then the $x^5$ coordinate($\equiv x$) 
of $NS5_R'$-brane is equal to
zero
and the $x^5$ coordinates of each half $NS5_L'$-brane are given by 
$\pm \Delta x$ respectively.
Giving an expectation value to the meson field $\Phi$
corresponds to recombination of $2N_c$- and $2N_c'$- color 
D4-branes, which will become $2N_c$-color D4-branes because 
$N_c'  > N_c$,
in Figure 1A such that they are suspended between 
the $NS5_L'$-brane and the $NS5_R'$-brane 
and pushing them into the $w \equiv x^8 + i
x^9$
direction. 
Now 
we draw this brane configuration in Figure 1B for nonvanishing mass
for the bifundamental $X$ by moving half of $NS5_L'$-brane with 
$N_c'$ color D4-branes to the $+v$ direction and their mirrors to $-v$
direction. 

\subsection{Magnetic theory}

By applying the Seiberg dual to the $Sp(N_c)$ factor in 
(\ref{gaugegroup}), the two $NS5_{L,R}'$-branes can be located at the
same side of the NS5-brane.
Starting from Figure 1B and moving the NS5-brane to the right all the
way past the $NS5_R'$-brane,
one obtains the Figure 2A.
Before arriving at the Figure 2A, there exists an intermediate 
step where the $N_c'$ D4-branes are connecting between half 
$NS5_L'$-brane and $NS5_R'$-brane(and their mirrors) and 
$2\widetilde{N}_c$ D4-branes connecting between $NS5_R'$-brane and   
NS5-brane. By introducing $2N_c'$ D4-branes and $2N_c'$ 
anti-D4-branes  between $NS5_R'$-brane and   
NS5-brane, reconnecting half of the former with  
the $N_c'$ D4-branes that are connecting between half 
$NS5_L'$-brane and $NS5_R'$-brane and moving those combined D4-branes
to $v$-direction(and their mirrors to $-v$ direction), 
one gets the final Figure 2A where we are left with 
$2(N_c'-\widetilde{N}_c)$ anti-D4-branes between $NS5_R'$-brane and   
NS5-brane.

\begin{figure}[ht]
   \epsfxsize=4.0in 
\centerline{\epsffile{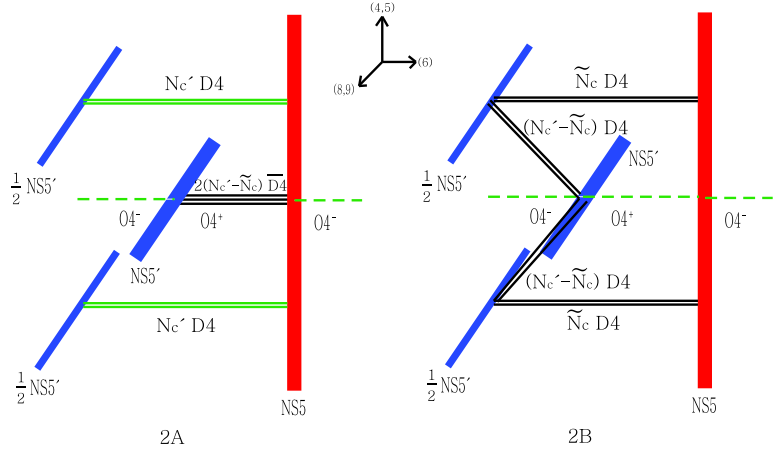}}
   \caption[FIG. \arabic{figure}.]{ 
The  magnetic brane configuration corresponding to Figure 1B with D4-
and $\overline{D4}$-branes(2A) and with 
a misalignment between D4-branes(2B) when the NS5'-branes are close to
each other.  }
\end{figure}

Then the gauge group is given by
\bea
Sp(\widetilde{N}_c=N_c'-N_c-2) \times SO(2N_c')
\label{dualgauge}
\eea
where the number of dual color is obtained from the linking number 
counting, as done in \cite{Ahn07-2}.
The matter contents are the field $Y$ in the bifundamental 
representation $({\bf 2\widetilde{N}_c, 2N_c'})$ under the dual gauge
group (\ref{dualgauge}) 
and  the gauge-singlet $\Phi$ for the first dual gauge group 
in the adjoint representation for 
the second dual gauge group, i.e.,
$({\bf 1, N_c'(2N_c'-1)})$ under the dual gauge group (\ref{dualgauge}).
These matter fields introduce 
a cubic superpotential which is an interaction between dual ``quarks''
$Y$ and a meson $\Phi$. 

Then the dual magnetic superpotential, by adding the mass term
(\ref{superelectric}) for the bifundamental $X$, which can be 
interpreted as a linear term in the meson $\Phi$, to this cubic
superpotential, is given by
\bea
W_{dual} =  \Phi Y Y + m \Phi. 
\label{superpo}
\eea
This can be seen from the equation (2.2) of \cite{Ahn07-2}
by removing the terms of D6-branes in electric and magnetic theories.
Of course, the brane configuration for zero mass for the bifundamental,
which has only a cubic superpotential,
can be obtained from Figure 2A by recombination between half
NS5'-branes together with color D4-branes via pushing 
them into the origin $v=0$.
Then the number of dual colors for D4-branes 
becomes $2N_c'$ between two NS5'-branes 
and $2\widetilde{N}_c$ between $NS5_R'$-brane and NS5-brane.
Or starting from Figure 1A and moving the NS5-brane to the right all the
way past the $NS5_R'$-brane,
one also obtains the corresponding magnetic brane configuration
for massless bifundamental.

The brane configuration in Figure 2A is stable as long as the
distance $\Delta x$ between the upper half $NS5_L'$-brane and
$NS5_R'$-brane is large, as
in \cite{GK}. If they are close to each other, then this brane
configuration is unstable to decay to the brane configuration in Figure
2B with bending effect of tilted 
D4-branes connecting half NS5'-brane and NS5'-brane.
One can regard these brane configurations as particular states in the
magnetic gauge theory with the gauge group (\ref{dualgauge}) and
superpotential (\ref{superpo}). The difference between the energies of
the configurations of Figure 2A and Figure 2B by evaluating the 
lengths of D4-branes in order to determine the true ground state 
can be obtained.
When the two half NS5'-branes are replaced by two coincident
D6-branes,
the brane configuration of Figure 2B is the same as the one studied in 
\cite{FGU,Ahn06-1}.

According to the result of \cite{GK}, the  flavor D4-branes of 
straight brane configuration
of
Figure 3 of \cite{GK}  can bend due to the fact that there exists an attractive
gravitational interaction
between those flavor D4-branes and NS5-brane from the DBI action. For
example, see the Figure 6 of \cite{GK} for explicit curve, obtained by
extremizing the DBI action, connecting
two NS5'-branes.
The correct phase transition of the classical string theory in the
background
and which one 
is the correct ground state of the system among the Figure 2 and 
the Figure 6 of \cite{GK} depends on the parameters $y_i$, the
locations of two NS5'-branes, and
$\Delta x$, the relative displacement between two NS5'-branes 
in the $v$-direction. 
For example, when the equal $y_i$ is less than a string scale
$\ell_s$,
for small $\Delta x$, the ground state is given by the brane configuration in
Figure 6 of \cite{GK} while for larger $\Delta x$, 
the ground state is given by 
the brane configuration in Figure 2 of \cite{GK}. 

One can perform similar analysis in our brane configuration with an
addition of O4-plane since one can take into account the behavior of
parameters geometrically in the presence of O4-plane.
Then the upper $(N_c'-\widetilde{N}_c)$ flavor D4-branes of 
straight brane configuration
of
Figure 2B can bend due to the fact that there exists an attractive
gravitational interaction
between those flavor D4-branes and NS5-brane from the DBI action, by
following the procedure of \cite{GK}. 
Of course, their mirrors, the lower 
$(N_c'-\widetilde{N}_c)$ flavor D4-branes of 
straight brane configuration
of
Figure 2B can bend and their trajectory connecting 
two NS5'-branes should be preserved under the O4-plane, i.e., ${\bf
Z}_2$ symmetric way, like as the symmetry property between the 
straight flavor D4-branes, when there is no gravitational interaction,
under the O4-plane in Figure 2B.
The correct choice for the ground state of the system 
depends on the parameters $y_i$ and $\Delta x$. 
When the equal $y_i$ is less than a string scale
$\ell_s$,
for small $\Delta x$, the ground state is the brane configuration in
Figure 2B with an appropriate bending effect and for larger $\Delta x$, 
the ground state is the brane configuration in Figure 2A. 

In the Figure 2B, the background geometry by NS5-brane gives
an attractive force between the upper tilted 
$(N_c'-\widetilde{N}_c=N_c+2)$ D4-branes and the NS5-brane. 
At first, we focus on the upper tilted D4-branes and later 
we'll describe its mirror, the lower tilted D4-branes in Figure 2B.
In order to
compute the upper bending curve for the D4-branes, i.e., a generalization of 
coincident straight lines due to an attractive force, 
connecting between the upper NS5'-brane and the middle
NS5'-brane in Figure 2B explicitly, 
one has to analyze the DBI action for
the upper tilted D4-branes 
in this background geometry. For example, see the
ref. \cite{Kutasov} for detailed explanations on brane dynamics near NS5-branes.

By following the procedure \cite{GK,Kutasov}, one can write down the DBI
action for the upper tilted D4-branes as, after
inserting both the dilaton and the induced metric,
\bea
S(upper) = - (N_c+2) \tau_4 \int d x \sqrt{\frac{1}{H(y)} +
  (\pa_x y)^2}, \qquad H(y) \equiv 1 +\frac{\ell_s^2}{y^2}  
\label{upper}
\eea
where $\tau_4$ is the tension of the D4-brane, the induced NS $B$
field vanishes \cite{Kutasov} 
and the harmonic function $H(y)$ is the field strength of NS $B$ field. In the
present case, the $l$ in $H(y)$ is equal to $\ell_s$ because we consider a
single NS5-brane background. 
The coordinate $r$ of \cite{GK} corresponds to the radial
coordinate away from the NS5-brane in the transverse (6789) 
directions, in general, but this will become $r=y=x^6$ at 
$x^7=x^8=x^9=0$. The constant term $+2$ in the number of D4-branes, 
$(N_c+2)$, appearing in front of
(\ref{upper}) 
is due to the presence
of O4-plane. This feature 
is a new fact, when we compare with the unitary case 
\cite{GK}. For the different O4-plane charge in the subsection 2.4,
we'll see the opposite coefficient for the constant term $-2$. 

Since the integrand of this action (\ref{upper}) does not depend on $x$ explicitly, 
there exists a conserved constant quantity \cite{GK}, through the Euler-Lagrange equation,
\bea
H(y) \sqrt{\frac{1}{H(y)} +
  (\pa_x y)^2}    = C = const.
\label{equation}
\eea
We are looking for a solution, by extremizing the DBI action (\ref{upper}), where 
the upper tilted D4-branes are described 
by a smooth coincident curve $y=y(x)$ connecting 
between the upper NS5'-brane and the middle NS5'-brane in Figure 2B.
Let us consider the solution of this equation of motion
(\ref{equation}) 
as a function
of a relative distance between two NS5'-branes
\bea
\Delta
x = x_2 -x_1=x_2
\nonu
\eea
where we put the $x$ coordinate of the upper
NS5'-brane in Figure 2B as $x=x_2$ while
the $x$ coordinate of the middle
NS5'-brane as $x=x_1=0$.
In other words, the O4-plane is located at $x=0$.
Then the mirror of the upper NS5'-brane, the lower NS5'-brane, 
is located at $x=-x_2$.
It is evident that for nonzero $\Delta x$ which corresponds to
massive case, 
the solution for bending
curve provides a
deformation of the  upper tilted $(N_c+2)$ coincident straight D4-branes. 
At the minimum value of $y$, which is denoted by $y_m$, 
the above equation (\ref{equation}) leads to
\bea
C^2 = H(y_m)
\label{C}
\eea 
since $\pa_x y$ at $y=y_m$ vanishes.

By using the separation of variables for $(x,y)$ in (\ref{equation})
together with (\ref{C}), 
one can write the integrals for the two intervals 
$0 \leq x \leq x_m$ and $x_m \leq x \leq x_2$ where the $x_m$ is the
corresponding 
$x$ coordinate on the curve to $y=y_m$. The former corresponds to 
the interval $y_1 \leq y \leq y_m$ while the latter does the interval 
$y_m \leq y \leq y_2$. We expect the same result
as the one in \cite{GK} because the brane configuration consisting of
two NS5'-branes, the NS5-brane and tilted D4-branes in Figure 2B is
exactly the same as the Figure 6 of \cite{GK} with an interchange of 
horizontal and perpendicular coordinates and the exact results for the
integrals turn out   
as follows \cite{GK}:
\bea
\int_{y_m}^{y_1} dy  \frac{H(y)}{\sqrt{H(y_m)-H(y)}} & = &
\frac{y_m}{\ell_s} \sqrt{y_1^2 -y_m^2} + \ell_s \theta_1  = - \int_{x_m}^{0}
d x = x_m,
\nonu \\
\int_{y_m}^{y_2} dy  \frac{H(y)}{\sqrt{H(y_m)-H(y)}} & = & 
\frac{y_m}{\ell_s} \sqrt{y_2^2 -y_m^2} + \ell_s \theta_2 = \int_{x_m}^{x_2}
d x = x_2-x_m
\label{integrals}
\eea
where we introduce the angles 
\bea
\cos \theta_i \equiv \frac{y_m}{y_i}, \qquad  
0 \leq \theta_i \leq \frac{\pi}{2},
\label{angle}
\eea
for $\theta_i \rightarrow 0$, the $y_m$ goes to $y_i$,
for $\theta_i \rightarrow \frac{\pi}{2}$, the $y_m$ approaches to zero,
and unfortunately, we use the same variables $y_i$ as 
the one in an electric theory and in Figure 1.
In other words, the $y$ coodinate for the upper NS5'-brane in Figure
2B is given by $y=y_2$ while 
the $y$ coodinate for the middle NS5'-brane in Figure
2B is given by $y=y_1$, as in \cite{GK}. The positive direction of $y$
in Figure 2B 
is directed to the left hand side of NS5-brane whose $y$ coordinate is zero. 
Note that the numerator $H(y)$ in the above integrands consists of two
parts, i.e., constant term and $y$-dependent term through (\ref{upper}). 
The constant term of numerator $H(y)$ in the integrals gives rise to the $1/\ell_s$
term in the middle of (\ref{integrals}) while
the $y$-dependent term of numerator $H(y)$ 
contributes to  the $\ell_s$ term in the middle of (\ref{integrals}), 
after the $y$-integrations. 

Now we have the explicit relation between $\Delta x$ and $y_i$ and $y_m$, 
by little algebra for the trigonometric functions, and by adding the two
integral results above (\ref{integrals}), as in \cite{GK},
the relative distance between two NS5'-branes for the curve, depending
on
$y_i$ and $y_m$, is given by
\bea
\Delta x(upper) = x_2 = \frac{1}{2 \ell_s} \left( y_1^2 \sin 2 \theta_1 +
  y_2^2 \sin 2 \theta_2 \right) + \ell_s \left( \theta_1 + \theta_2 \right).
\label{distance}
\eea
This is invariant under $y_i \rightarrow -y_i$ and $y_m \rightarrow -y_m$. 
Note that $\theta_i$ is also invariant under these transformations, 
by (\ref{angle}).
When $\theta_i=0$, then $\Delta x =0$ which is for the massless case 
and for $\theta_i
=\frac{\pi}{2}$,
the $\Delta x$ is equal to $\pi \ell_s$ that corresponds to 
the configuration of
Figure 2A. 

Moreover, 
the energy of the Figure 2B with bending effect for the upper
D4-branes is given by 
\bea
E_{curved}(upper) = -S(upper)
\label{curupper}
\eea 
together with (\ref{upper}).
By using the relation 
(\ref{C}) and change the integration over $x$ into the $y$ variable
with (\ref{equation}), this leads to $E_{curved}(upper) =(N_c+2) \tau_4
\sqrt{H(y_m)} 
\int dy  
\frac{1}{\sqrt{H(y_m)-H(y)}} $. Then 
one arrives at the following expression for the energy for D4-branes, 
as in \cite{GK}, 
by adding the results of 
(\ref{integrals}) that contain 
$1/\ell_s$ term, as we mentioned before,
\bea
E_{curved}(upper) = (N_c+2) \tau_4
\frac{\sqrt{H(y_m)}}{2\ell_s}
\left( y_1^2 \sin 2 \theta_1 +
  y_2^2 \sin 2 \theta_2 \right).
\label{energy}
\eea
In other words, the first two terms in (\ref{distance}) appear in 
(\ref{energy}) as a factor. 
This energy (\ref{energy}) is also
invariant under $y_i \rightarrow -y_i$ and $y_m \rightarrow -y_m$ with
(\ref{angle}). 
Note that there exists a constant term $+2$ coming from the O4-plane, 
in the overall coefficient of (\ref{energy}).

So far, we have only considered the contributions, i.e., the explicit
curve connecting two NS5'-branes and the energy of the configuration
in that background, from the bending
effect of upper tilted D4-branes
in Figure 2B. Now we can compute the contributions from their mirrors,
i.e., the lower tilted $(N_c+2)$ D4-branes in Figure 2B. 
The DBI action for these D4-branes, $S(lower)$, is 
the same as above $S(upper)$ given in (\ref{upper}) 
since the number of D4-branes are the
same and the background geometry is characterized by the same
NS5-brane, implying that the induced metric and a dilaton are 
the same as before. That is, 
\bea
S(lower)=S(upper)
\label{lowerupper1}
\eea 
with (\ref{upper}). 
It is straightfoward to see that there exist a conserved quantity 
(\ref{equation}) and a relation (\ref{C}).
Now the coordinate $x$ of lower NS5'-brane in Figure 2B is given 
by $x=-x_2$, as mentioned before. 
Remember that the O4-plane action restricts the position
of lower NS5'-brane in this particular way. Of course, 
the coordinate $x$ of middle NS5'-brane is equal to $x=0$.
Also the $y$ coordinate of lower NS5'-brane in Figure 2B 
is the same as 
$y=y_2$ for the upper NS5'-brane.

By using the separation of variables from (\ref{equation}) with (\ref{C}), 
one can write the integrals for the two intervals 
$-x_m \leq x \leq 0$ and $-x_2 \leq x \leq -x_m$, as we did before.
Note that the $x$ values for the bending curve are all negative except
the one of the middle NS5'-brane. 
It turns out the results are given by (\ref{integrals}) and
(\ref{angle}) again. 
Moreover the relative distance $\Delta x (lower) = 0-(-x_2)
=x_2$ between the middle NS5'-brane and the lower NS5'-brane 
is given by (\ref{distance}):
\bea
\Delta x (lower) = \Delta x (upper).
\nonu
\eea
The ${\bf Z}_2$ symmetry by an O4-plane acting as $(x,y) \rightarrow
(-x,y)$ 
reflects here if we take the absolute value for $\Delta x$.
The two 
bending curves for the upper and lower tilted D4-branes are symmetric
each other under O4-plane.
Since the energy of the Figure 2B with bending effect for the lower
tilted D4-branes is given by 
minus $S(lower)$ which is related 
to (\ref{lowerupper1}) and further (\ref{curupper}), 
eventually one obtains that 
\bea
E_{curved}(lower)=
E_{curved}(upper)
\nonu
\eea 
with (\ref{energy}).

Then the analysis of \cite{GK} can be done, using the results of
both (\ref{distance}) which is exactly the same as the one in \cite{GK}
and (\ref{energy}) which has different overall coefficient due to the
O4-plane when we compare with the unitary case \cite{GK}, 
for equal $y_i$'s and unequal $y_i$'s.
Since the correct choice for the ground state of the system 
depends on the parameters $y_i$ and $\Delta x$, once we understand the
right phase structure for the upper tilted case given by
(\ref{distance}) 
and (\ref{energy}), then the corresponding phase structure for its mirror, the
lower tilted case,  
is satisfied automatically. 

\subsection{Gauge theory analysis at small $\Delta x$, a 
mass for the fundamental }

The quantum corrections can be understood for small $\Delta x$ by 
using the low energy field theory on the branes.
The low energy dynamics of the magnetic brane configuration 
can be described by the ${\cal N}=1$ supersymmetric gauge theory
with gauge group (\ref{dualgauge})
and the gauge couplings for the two gauge group factors are
given by
\bea
g_{Sp,mag}^2 = \frac{g_s \ell_s}{y_2}, \qquad 
g_{SO,mag}^2 = \frac{g_s \ell_s}{(y_1-y_2)}.
\label{coupling}
\eea
In the classical string theory, the gauge theory is weakly coupled,
i.e.,
small $g_s$ with fixed $\frac{y_i}{\ell_s}$.
By tuning $y_1$ and $y_2$, one of the gauge couplings can be larger
than the other. 

The dual gauge theory has  an adjoint $\Phi$ of $SO(2N_c')$, i.e., an
antisymmetric matrix and 
bifundamental $Y$ in the representation 
 $({\bf 2\widetilde{N}_c, 2N_c'})$ under the dual gauge
group (\ref{dualgauge}) and the superpotential 
corresponding to Figures 2A and 2B is given by 
\bea
W_{dual} = h \Phi Y Y  - h \mu^2 \Phi, \qquad
h^2=  g_{SO,mag}^2 
\nonu
\eea
in the parametrization of \cite{ISS} and we used the equation of (2.9)
of \cite{BGHSS} for the value of $h$.
Here the mass parameter is given by
\bea
\mu^2 = -\frac{\Delta x}{2\pi g_s \ell_s^3}. 
\label{mass}
\eea
That is, the 
second term in the superpotential measures the separation of 
the NS5'-branes in the $x$ direction.
Then $ Y Y$ is a $2\widetilde{N}_c \times 2\widetilde{N}_c$ 
matrix where the second gauge group indices for two $Y$'s 
are contracted with those
of $\Phi$ while $\mu^2$ is a 
$2N_c' \times 2N_c'$ antisymmetric matrix.
Although the field $Y$ itself is a fundamental in the second gauge
group
which is a different feature, compared with the singlet 
representation for the usual quark
coming from D6-branes \cite{Ahn07-2},
the product $Y Y$ has the same representation with the 
product of quarks, $\widetilde{Q} \widetilde{Q}$ in the notation of 
\cite{Ahn07-2}.
Moreover, the second gauge group indices for the field $\Phi$ play the
role of the flavor indices for the gauge singlet $S \equiv
Q Q $ in \cite{Ahn07-2}.

Therefore, the F-term equation, the derivative $W_{dual}$ with respect to the
meson field $\Phi$ cannot be satisfied if the $2N_c'$ exceeds
$2\widetilde{N}_c$.
So the supersymmetry is broken.   
That is, 
there are two equations from F-term conditions:
$
YY -\mu^2 =0$ and $ \Phi Y =0$.
Then the solutions for these
are given by 
\bea
<Y>   = 
\left(
\begin{array}{c}
\mu  {\bf 1}_{2\widetilde{N}_c}  \\
0
\end{array}
\right), 
\qquad
<\Phi> =
 \left(
\begin{array}{cc}
0  & 0  \\
0 & \Phi_0  {\bf 1}_{(N_c'-\widetilde{N}_c)} \otimes i \sigma_2
\end{array}
\right). 
\label{point}
\eea
Then one can expand these fields around on a point (\ref{point}), as
in \cite{ISS,Ahn07-2} and one arrives at the relevant superpotential
up to quadratic order in the fluctuation. 
At one loop, the effective potential $V_{eff}^{(1)}$ for $\Phi_0$
leads to the positive value for $m_{\Phi_0}^2$ implying that these
vacua are stable.

By extremizing the low energy superpotential \cite{ISS}, 
the supersymmetric vacua occur at
\bea
<h \Phi> = \Lambda_1 \left(\frac{\mu}{\Lambda_1}\right)^{
\frac{2(\widetilde{N}_c+1)}{(N_c'-\widetilde{N}_c-1)}} {\bf 1}_{N_c'}
\otimes i \sigma_2.
\label{phi}
\eea
We are interested in the case where $N_c' > 3(\widetilde{N}_c+1)$ so
that $Sp(\widetilde{N}_c)$ gauge coupling is IR free. It becomes 
strongly coupled at the scale
\bea
\Lambda_1 = E_c \exp \left[ \frac{8\pi^2 y_2}{(N_c'-3(\widetilde{N}_c+1))
    g_s \ell_s } \right]
\label{scale}
\eea
where the expression (\ref{coupling}) is used.
Then, the condition that
$<h \Phi> $ is much smaller than  $E_c$
implies, by plugging (\ref{mass}) and (\ref{scale}) into (\ref{phi}),  
that the gauge theory analysis is only valid in the regime where 
$\Delta x$ is smaller than $\exp(-\frac{C}{g_s})$ with some positive
constant $C$ \cite{GK}. 

\subsection{Other magnetic theory with same electric theory}

By applying the Seiberg dual to the $SO(2N_c')$ factor in 
(\ref{gaugegroup}), the two $NS5_{L,R}'$-branes can be located at the
right side of the NS5-brane.
Starting from modified Figure 1B, where 
the $x^5$ coordinate 
of $NS5_L'$-brane is equal to
zero
and the $x^5$ coordinates of half $NS5_R'$-brane are $\pm \Delta x$,
and moving the NS5-brane to the left all the
way past the $NS5_L'$-brane,
one obtains the magnetic brane configuration similar to Figure 2A.
The gauge group is given by
\bea
Sp(N_c) 
\times SO(2\widetilde{N}_c'=2N_c-2N_c'+4).
\label{gaugedual}
\eea
The matter contents are the field $Y$ in the bifundamental 
representation $({\bf 2N_c, 2\widetilde{N}_c'})$ under the dual gauge
group (\ref{gaugedual}) 
and  the gauge-singlet $\Phi$ for the second dual gauge group 
in the adjoint representation for 
the first dual gauge group, i.e., a symmetric matrix,
$({\bf N_c(2N_c+1), 1})$ under the dual gauge group.
The superpotential is the same as the one in (\ref{superpo}) and the
corresponding Figure 2B, which is exactly a reflection of Figure 2B with
respect to the NS5-brane, i.e., all the D4-branes and NS5'-branes are
located at the right hand side of NS5-brane, 
can be constructed similarly.
The DBI analysis done in previous case can be obtained also in this case.
The number of relevant D4-branes here is given by $(N_c'-2)$ which
plays the role of $(N_c+2)$ in previous section.

The gauge couplings for the two gauge group factors are
given by
\bea
g_{Sp,mag}^2 = \frac{g_s \ell_s}{(y_2-y_1)}, \qquad 
g_{SO,mag}^2 = \frac{g_s \ell_s}{y_1}
\nonu
\eea
and
the superpotential 
corresponding to modified Figures 2A and 2B is given by 
\bea
W_{dual} = h \Phi Y Y  - h \mu^2 \Phi, \qquad
h^2=  g_{Sp,mag}^2 
\nonu
\eea
where the  mass parameter $\mu^2$ is given by (\ref{mass}).
Then the solutions for these
are given by 
\bea
<Y>   = 
\left(
\begin{array}{c}
\mu  {\bf 1}_{2\widetilde{N}_c'}  \nonu \\
0
\end{array}
\right), 
\qquad
<\Phi> =
 \left(
\begin{array}{cc}
0  & 0
 \\
0 & \Phi_0  {\bf 1}_{2(N_c-\widetilde{N}_c')} 
\end{array}
\right).
\nonu
\eea
At one loop, the effective potential $V_{eff}^{(1)}$ for $\Phi_0$
leads to the positive value for $m_{\Phi_0}^2$ implying that these
vacua are stable.

By extremizing the low energy superpotential \cite{ISS}, 
the supersymmetric vacua occur at
\bea
<h \Phi> = \Lambda_2 \left(\frac{\mu}{\Lambda_2}\right)^{
\frac{2(\widetilde{N}_c'-2)}{(N_c-\widetilde{N}_c'+2)}} {\bf 1}_{N_c}.
\nonu
\eea
We are interested in the case where $N_c > 3(\widetilde{N}_c'-2)$ so
that $SO(2\widetilde{N}_c')$ gauge coupling is IR free. It becomes 
strongly coupled at the scale
\bea
\Lambda_2 = E_c \exp \left[ \frac{8\pi^2 y_1}{(N_c-3(\widetilde{N}_c'-2))
    g_s \ell_s } \right].
\nonu
\eea
Then, the condition that
$<h \Phi> $ is much smaller than  $E_c$
implies  
that the gauge theory analysis is only valid in the regime where 
$\Delta x$ is smaller than $\exp(-\frac{C}{g_s})$ with some positive
constant $C$. 

\section{When an O6-plane is added}

In this section, we add an orientifold 6-plane to the type IIA brane
configurations of \cite{GK} together with two extra outer NS5-branes 
and construct new meta-stable brane configurations. 
For the second example, we add D6-branes more.

\subsection{Electric theory}

The type IIA brane configuration corresponding to 
${\cal N}=1$ supersymmetric gauge theory with
gauge group
\bea
SU(N_c) \times SU(N_c')
\label{secondgauge}
\eea
and the symmetric flavor for $SU(N_c)$, the conjugate 
symmetric flavor for $SU(N_c)$,
a bifundamental $X$ in the representation 
$({\bf N_c, \overline{N_c'}})$ and its conjugate field $\widetilde{X}$ 
in the representation $({\bf \overline{N_c}, N_c'})$, 
under the gauge group can be described similarly. 
It consists of 
a middle $NS5_M$-brane(012345),
the left $NS5_L$-brane(012345) and the right 
$NS5_R$-brane(012345), the left $NS5_L'$-brane(012389) and the right 
$NS5_R'$-brane(012389), $N_c$-  and $N_c'$-color D4-branes(01236) and an 
$O6^{+}$-plane(0123789).
We take the arbitrary number of color D4-branes with the constraint 
$2N_c' \geq N_c$.
The $O6^{+}$-plane acts as $(x^4,x^5,x^6) \rightarrow
(-x^4,-x^5,-x^6)$ and has RR charge $+ 4$ 
playing the role of $+4$
D6-brane.
The bifundamentals $X$ and $\widetilde{X}$  correspond to 4-4 
strings connecting 
the $N_c$-color D4-branes with $N_c'$-color D4-branes.
The symmetric and conjugate symmetric flavors correspond to 
4-4 strings connecting $N_c$ D4-branes located at negative $x^6$
region
and $N_c$ D4-branes located at positive $x^6$ region.
See also the relevant works in \cite{ILS,BH,BIWW,LLL}.

The middle NS5-brane is located at $x^6=0$ and the $x^6$ 
coordinates for the $NS5_L$-brane, $NS5_L'$-brane, $NS5_R'$-brane and
$NS5_R$-brane are given  by $x^6=-y_2, -y_1, y_1$ and
$x^6=y_2$
respectively, along the line of \cite{GK}. 
The $N_c$ D4-branes 
are suspended between the 
$NS5_L'$-brane, whose $x^6$ coordinate is given by $x^6=-y_1$,  and 
$NS5_R'$-brane, whose $x^6$ coordinate is given by $x^6=y_1$,  while 
the $N_c'$ D4-branes 
are suspended between the $NS5_L$-brane and the $NS5_L'$-brane
and moreover they are 
suspended between the $NS5_R'$-brane and the $NS5_R$-brane.
We draw this brane configuration in Figure 3A for the vanishing mass
for the bifundamentals. See also 
the relevant previous work appeared in 
\cite{Ahn07-4}
\footnote{This is equivalent to the reduced brane
  configuration of \cite{Ahn07-4} with particular rotations for the
  NS5-branes if we remove all the D6-branes 
completely.}.

\begin{figure}[ht]
   \epsfxsize=5.0in 
\centerline{\epsffile{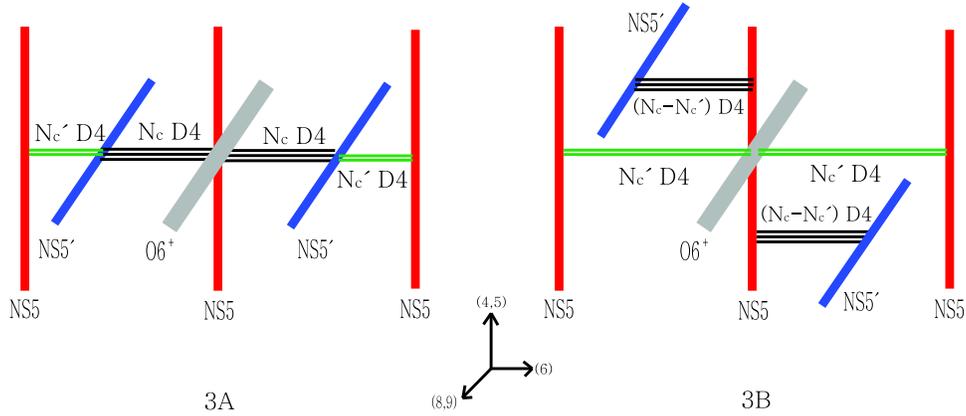}}
   \caption[FIG. \arabic{figure}.]{ 
The  ${\cal N}=1$ supersymmetric 
electric brane configuration for the gauge group $SU(N_c) \times
SU(N_c')$ and the bifundamentals $X$ and $\widetilde{X}$  
as well as symmetric and conjugate symmetric flavors with 
vanishing(3A) and 
nonvanishing(3B) mass
for the bifundamentals corresponding to the dual for the second gauge
group. 
The bifundamentals $X$ and $\widetilde{X}$  
correspond to 4-4 
strings connecting 
the $N_c$-color D4-branes with $N_c'$-color D4-branes. }
\end{figure}

The gauge couplings of $SU(N_c)$ and $ SU(N_c')$
are given by
\bea
g_1^2 = \frac{g_s \ell_s}{y_1}, \qquad 
g_2^2 = \frac{g_s \ell_s}{y_2}.
\label{couplings}
\eea
As $y_2$ goes to $\infty$, the $SU(N_c')$ gauge group becomes a
global symmetry and the theory leads to SQCD-like theory with the gauge group
$SU(N_c)$ with symmetric and conjugate symmetric flavors 
and $N_c'$ flavors in the fundamental representation.

According to result of \cite{Ahn07-4}, there is no electric
superpotential corresponding to the Figure 3A. Now let us deform 
this theory. 
Displacing the two NS5'-branes relative each other in the $v$ 
direction corresponds to turning on a quadratic
superpotential
for the bifundamentals $X$ and $\widetilde{X}$ as follows:
\bea
W = m X \widetilde{X} \left( \equiv m \Phi' \right)
\label{superelectric1}
\eea
where the $\Phi'$
is a meson field
and the mass $m$ is given by geometrically
\bea
m = \frac{\Delta x}{2 \pi \alpha'} 
\left( = \frac{\Delta x}{\ell_s^2} \right).
\label{mass1}
\eea
The $NS5_L'$-brane is moving to the $+v$ direction and the 
$NS5_R'$-brane is moving to $-v$ direction due to the O6-plane for
fixed NS5-branes.
That is, the $x^5$ coordinate of $NS5_L'$-brane is $+ \Delta x$ while
the $x^5$ coordinate of $NS5_R'$-brane is $- \Delta x$.
We draw this brane configuration in Figure 3B for nonvanishing mass
for the bifundamentals
by moving the $NS5_L'$-brane with 
$(N_c-N_c')$ color D4-branes 
to the $+v$ direction and their mirrors to $-v$
direction. 
For the meta-stable brane configuration next subsection, we need to
move outer NS5-branes rather than NS5'-branes.  

\subsection{Magnetic theory}

Let us consider two separate cases.

$\bullet$ When the dual magnetic case is taken from the second gauge group

By applying the Seiberg dual to the $SU(N_c')$ factor in 
(\ref{secondgauge}), the two $NS5_{L,R}'$-branes can be located at the
outside of the three NS5-branes.
Starting from Figure 3B and moving the $NS5_R'$-brane to the right all the
way past the $NS5_R$-brane and then taking $\frac{\pi}{2}$ 
rotations of two outer NS5-branes,
there exist the $\widetilde{N}_c'(=N_c-N_c')$ 
D4-branes that are connecting between two 
NS5'-brane(and their mirrors) and 
$N_c$ D4-branes connecting between NS5'-brane and   
NS5-brane. 
Since $\widetilde{N}_c'$ is less than $N_c$, it is not possible to 
construct a misalignment of the flavor D4-branes. 
Therefore, there is no meta-stable brane configuration in this case.

$\bullet$ When the dual magnetic case is taken from the first gauge group

Starting from the Figure 3A, we apply the Seiberg dual to the 
$SU(N_c)$ factor in (\ref{secondgauge}), the two NS5'-branes are 
interchanged each other. Then the number of color $\widetilde{N}_c$
is given by $\widetilde{N}_c=2N_c'-N_c$ from \cite{Ahn07-4,Ahn07}.
By rotating the outer two NS5-branes by $\frac{\pi}{2}$ and moving
them
to $\pm v$ direction, 
the $N_c'$ 
D4-branes are connecting between two 
NS5'-branes(and their mirrors) and 
$\widetilde{N}_c$ D4-branes connecting between NS5'-brane and   
NS5-brane. 
By introducing $N_c'$ D4-branes and $N_c'$ 
anti-D4-branes  between NS5'-brane and   
NS5-brane, reconnecting the former with  
the $N_c'$ D4-branes connecting between two NS5'-branes 
and moving those combined D4-branes
to $v$-direction(and their mirrors to $-v$ direction), 
one gets the final Figure 4A where we are left with 
$(N_c'-\widetilde{N}_c)$ anti-D4-branes between NS5'-brane and   
NS5-brane.

\begin{figure}[ht]
   \epsfxsize=5.0in 
\centerline{\epsffile{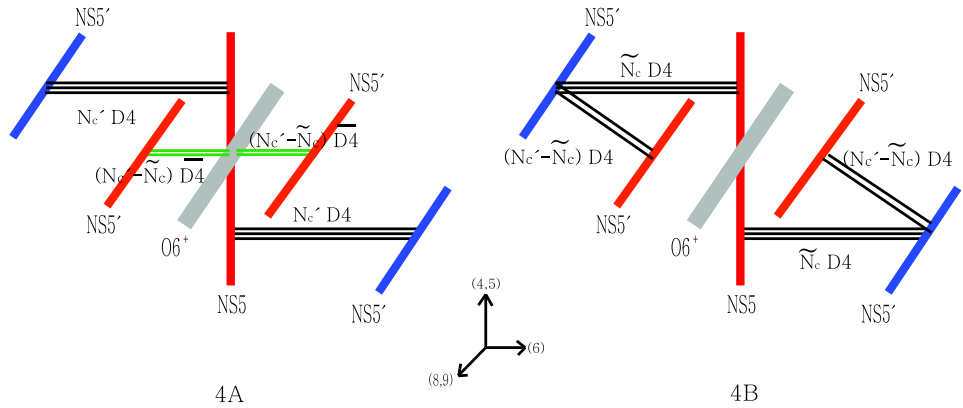}}
   \caption[FIG. \arabic{figure}.]{ 
The  magnetic brane configuration corresponding to Figure 3A with D4-
and $\overline{D4}$-branes(4A) and  with 
a misalignment between D4-branes(4B)  when the NS5'-branes are close to
each other. }
\end{figure}

The gauge group is given by
\bea
SU(\widetilde{N}_c=2N_c'-N_c) \times SU(N_c')
\label{dual2}
\eea
where the number of dual color can be obtained from the linking number 
counting, as done in \cite{Ahn07-4,Ahn07}.
The matter contents are the flavor singlet $Y$ in the bifundamental 
representation $({\bf \widetilde{N}_c, \overline{N_c'}})$
and its complex conjugate field $\widetilde{Y}$ in the bifundamental 
representation  $({\bf \overline{\widetilde{N}_c}, N_c'})$,
under the dual gauge
group (\ref{dual2}) 
and  the gauge singlet $\Phi'$ in the representation for 
$({\bf 1,N_c^{'2}-1}) \oplus ({\bf 1, 1})$ under the dual gauge group.
There are also
the symmetric flavor for $SU(\widetilde{N}_c)$ and the conjugate 
symmetric flavor for $SU(\widetilde{N}_c)$.
A cubic superpotential is an interaction between dual ``quarks''
and a meson. 

Then the dual magnetic superpotential, by adding the mass term
like as (\ref{superelectric1}) for the bifundamental $X$ which can be 
interpreted as a linear term in the meson $\Phi'$ to this cubic
superpotential, is given by
\bea
W_{dual} =  \Phi' Y \widetilde{Y} + m \Phi' 
\label{superpo1}
\eea
where this can be seen from the 
equation (2.3) of \cite{Ahn07-4} by putting the terms coming from  
the D6-branes in both electric and magnetic theories to zero.
The brane configuration for zero mass for the bifundamentals
can be obtained from Figure 4A by  pushing the two NS5'-branes  
into the origin $v=0$.
Then the number of dual colors for D4-branes 
becomes $N_c'$ between two NS5'-branes
and $\widetilde{N}_c$ between the NS5'-brane and the NS5-brane.

The brane configuration in Figure 4A is stable as long as the
distance $\Delta x$ between the upper $NS5'$-brane and 
the middle $NS5'$-brane is large, as
in \cite{GK}. If they are close to each other, then this brane
configuration is unstable to decay and leads to 
the brane configuration in Figure
4B.
One can regard these brane configurations as particular states in the
magnetic gauge theory with the gauge group (\ref{dual2}) and
superpotential (\ref{superpo1}).
When the two NS5'-branes which are connected by $\widetilde{N}_c$
D4-branes 
are replaced by two coincident
D6-branes,
the brane configuration of Figure 4B is the same as the one studied in 
\cite{Ahn07,FGU}.

One can perform similar analysis in our brane configuration 
since one can take into account the behavior of
parameters geometrically in the presence of O6-plane.
Then the upper $(N_c'-\widetilde{N}_c)$ flavor D4-branes of 
straight brane configuration
of
Figure 4B can bend due to the fact that there exists an attractive
gravitational interaction
between those flavor D4-branes and NS5-brane from the DBI action, by
following the procedure of \cite{GK}. 
Of course, their mirrors, the lower 
$(N_c'-\widetilde{N}_c)$ flavor D4-branes of 
straight brane configuration
of
Figure 4B can bend and their trajectory connecting 
two NS5'-branes should be preserved under the O6-plane, i.e., ${\bf
Z}_2$ symmetric way, like as the symmetry property between the 
straight flavor D4-branes, when there is no gravitational interaction,
under the O6-plane in Figure 4B.
The correct choice for the ground state of the system 
depends on the parameters $y_i$ and $\Delta x$. 

In the remaining paragraphs, we describe DBI action very briefly 
since the main discussions are done in previous subsection 2.2.
Since the brane geometry of the upper tilted
$(N_c'-\widetilde{N}_c=N_c-N_c')$ 
D4-branes, two
NS5'-branes 
and NS5-brane 
in Figure 4B 
is exactly the same as the one in Figure 2B except that the
corresponding number of
D4-branes is different, 
the DBI analysis can be done straightforwardly by following the
previous procedure.
One can write down the DBI
action as \cite{GK}
\bea
S(upper) = - (N_c-N_c') \tau_4 \int d x \sqrt{\frac{1}{H(y)} +
  (\pa_x y)^2}, \qquad H(y) \equiv 1 +\frac{\ell_s^2}{y^2}  
\label{upper1}
\eea
where we inserted the correct number of D4-branes.
The presence of $N_c'$-dependence(as well as $N_c$-dependence) 
in front of (\ref{upper1})
comes from the fact that the number of dual colors 
depends on how one takes the Seiberg dual strictly \cite{Ahn07}.
This is a different aspect, compared with the one in \cite{GK} or
the previous case considered in subsection 2.2.
In this case also, there are a conserved quantity (\ref{equation}) and
a relation by (\ref{C}) because the number of D4-branes does not
depend on these equations. 
The relative distance between the two NS5'-branes is characterized by 
(\ref{distance}). We rewrite here for convenience
\bea
\Delta x(upper) = x_2 = \frac{1}{2 \ell_s} \left( y_1^2 \sin 2 \theta_1 +
  y_2^2 \sin 2 \theta_2 \right) + \ell_s \left( \theta_1 + \theta_2 \right).
\label{distance1}
\eea
Finally, the energy \cite{GK} of the Figure 4B with bending effect, coming from
the upper tilted D4-branes, is written as 
\bea
E_{curved}(upper) = (N_c-N_c') \tau_4
\frac{\sqrt{H(y_m)}}{2\ell_s}
\left( y_1^2 \sin 2 \theta_1 +
  y_2^2 \sin 2 \theta_2 \right)
\label{energy1}
\eea
where we also put the correct number of D4-branes for the present case
and all the
variables $(y_i, y_m, \theta_i)$ are the same as the one in subsection 2.2.

So far, we have only considered the contributions from the bending
effect of upper tilted D4-branes
in Figure 4B. Now we can compute the contributions from their mirrors,
i.e., the lower tilted $(N_c-N_c')$ D4-branes in Figure 4B. 
The DBI action for these D4-branes, $S(lower)$, is 
the same as above $S(upper)$ since the number of D4-branes are the
same and the background geometry is characterized by the same
NS5-brane implying that the induced metric and dilaton are 
the same as before. That is, 
\bea
S(lower)=S(upper)
\label{lowerupper}
\eea 
with (\ref{upper1}). 
It is straightfoward to see that there exist also a conserved quantity 
(\ref{equation}) and a relation (\ref{C}). Note that 
under the replacement $y_m$ by $-y_m$ which is the maximum value of
$y$(note that the positive direction for $y$ is directed to the left
hand side of the NS5-brane in Figure 4B), 
this relation (\ref{C}) still holds.
Now the coordinate $x$ of lower NS5'-brane in Figure 4B is given 
by $x=-x_2$ while the coordinate $y$ of it is given by $y=-y_2$.
Recall that the O6-plane action reflects here also, as in Figure 4B. 
One can view the brane configuration consisting of 
the lower NS5'-branes and D4-branes in Figure 4B as 
the brane configuration after taking a reflection for lower
NS5'-branes and D4-branes in Figure 2B with respect
to the NS5-brane($y=0$ plane). This procedure 
is equivalent to transform the $y_i$ 
coordinates for NS5'-branes in Figure 2B as $-y_i$ keeping $x$ coordinates unchanged. 
Or equivalently,  
the two 
bending curves for the upper and lower tilted D4-branes in Figure 4B are symmetric
each other under the origin $(x,y)=(0,0)$.

By using the separation of variables in (\ref{equation}), 
one can write the integrals for the two intervals 
$-x_m \leq x \leq 0$ and $-x_2 \leq x \leq -x_m$, as before.
It turns out the results are given by (\ref{integrals}) and
(\ref{angle}) which is invariant under the 
$y_i \rightarrow -y_i$ and $y_m \rightarrow -y_m$. 
Moreover, the relative distance between two NS5'-branes, 
$\Delta x (lower) = 0-(-x_2)
=x_2$, is given by (\ref{distance1}):
\bea
\Delta x (lower) = \Delta x (upper).
\nonu
\eea
Since the energy of the Figure 4B with bending effect for the lower
tilted D4-branes is given by 
minus $S(lower)$ together with (\ref{lowerupper}), 
one obtains that 
\bea
E_{curved}(lower)=E_{curved}(upper)
\nonu
\eea 
with (\ref{energy1}).
Note that although the functions $\Delta x(upper)$ and $E_{curved}(upper)$
depend on $y_i$ and $y_m$, the replacements $y_i \rightarrow -y_i$ and
$y_m \rightarrow -y_m$ does not change these functions.
The ${\bf Z}_2$ symmetry by an O6-plane acting as $(x,y) \rightarrow
(-x,-y)$ 
reflects here.  

Then the analysis of \cite{GK} can be done using the results of
(\ref{distance1}) which is exactly the same as the one in \cite{GK}
and (\ref{energy1}) which has different overall coefficient containing
the rank of the second gauge group $N_c'$ due to the
O6-plane when we compare with the unitary case, 
for equal $y_i$'s and unequal $y_i$'s.
Once we understand the
correct phase structure for the upper tilted D4-branes case which will
be the same as the one \cite{GK} basically, then 
the corresponding analysis for its mirror, 
lower tilted D4-branes case is satisfied automatically. 

\subsection{Gauge theory analysis at small $\Delta x$ }

The low energy dynamics of the magnetic brane configuration 
can be described by the ${\cal N}=1$ supersymmetric gauge theory
with gauge group (\ref{dual2})
and the gauge couplings for the two gauge group factors are
given by
\bea
g_{1,mag}^2 = \frac{g_s \ell_s}{y_1}, \qquad 
g_{2,mag}^2 = \frac{g_s \ell_s}{y_2-y_1}.
\nonu
\eea
The dual gauge theory has  an adjoint $\Phi'$ of $SU(N_c')$ and 
bifundamental $Y$ in the representation 
 $({\bf \widetilde{N}_c, \overline{N_c'}})$ under the dual gauge
group (\ref{dual2}) and the superpotential 
corresponding to Figures 4A and 4B is given by
\bea
W_{dual} = h \Phi' Y \widetilde{Y} - h \mu^2 \Phi', \qquad
h^2=  g_{2,mag}^2 
\nonu
\eea
and the  mass parameter $\mu^2$ is given by (\ref{mass}).
Then $ Y \widetilde{Y}$ is 
a $\widetilde{N}_c \times \widetilde{N}_c$ 
matrix where the second gauge group indices for $Y$ and $\widetilde{Y}$ 
are contracted with those
of $\Phi'$ while $\mu^2$ is a 
$N_c' \times N_c'$ matrix.
Although the field $Y$ itself is a fundamental in the second gauge
group
which is a different feature, compared with the singlet 
representation for the usual quark
coming from D6-branes \cite{Ahn07-4},
the product $Y \widetilde{Y}$ has the same representation with the 
product, $q \widetilde{s} s \widetilde{q}$ in the notation of 
\cite{Ahn07-4}.
Moreover, the second gauge group indices for the field $\Phi'$ play the
role of the flavor indices for the gauge singlet $M' \equiv
Q \widetilde{Q} $ in \cite{Ahn07-4}.

Therefore, the F-term equation, the derivative $W_{dual}$ with respect to the
meson field $\Phi'$ cannot be satisfied if the $N_c'$ exceeds
$\widetilde{N}_c$.
So the supersymmetry is broken.   
That is, 
there are three equations from F-term conditions:
$
Y \widetilde{Y} -\mu^2 =0, \Phi' Y =0$, and $\widetilde{Y} \Phi'=0$.
Then the solutions for these
are given by 
\bea
<Y>   = 
\left(
\begin{array}{c}
\mu  e^{\phi} {\bf 1}_{\widetilde{N}_c}  \nonu \\
0
\end{array}
\right), \quad
<\widetilde{Y}>   = 
\left(
\begin{array}{cc}
\mu  e^{-\phi} {\bf 1}_{\widetilde{N}_c} & 0 \nonu \\
\end{array}
\right), 
\quad
<\Phi'> =
 \left(
\begin{array}{cc}
0  & 0
 \\
0 & \Phi_0  {\bf 1}_{(N_c'-\widetilde{N}_c)} 
\end{array}
\right).
\nonu
\eea
At one loop, the effective potential $V_{eff}^{(1)}$ for $\Phi_0$
leads to the positive value for $m_{\Phi_0}^2$ implying that these
vacua are stable.
The gauge theory analysis where the theory will be strongly coupled in
the IR region $N_c' > 2\widetilde{N}_c-2$ is only valid in the regime where 
$\Delta x$ is smaller than $\exp(-\frac{C}{g_s})$ with some positive
constant $C$. 

\subsection{Other electric and magnetic theories with same gauge group
  and different matters}


The type IIA brane configuration corresponding to 
${\cal N}=1$ supersymmetric gauge theory with
gauge group (\ref{secondgauge})
and the antisymmetric flavor for $SU(N_c)$, the conjugate 
symmetric flavor for $SU(N_c)$, eight fundamentals for $SU(N_c)$,
a bifundamental $X$ in the representation 
$({\bf N_c, \overline{N_c'}})$ and its conjugate field $\widetilde{X}$ 
in the representation $({\bf \overline{N_c}, N_c'})$, 
under the gauge group can be described similarly. 
It consists of 
a middle $NS5_M'$-brane,
the left $NS5_L'$-brane and the right 
$NS5_R'$-brane, the left $NS5_L$-brane and the right 
$NS5_R$-brane, $N_c$-  and $N_c'$-color D4-branes,
eight semi-infinite D6-branes, an 
$O6^{+}$-plane and $O6^{-}$-plane.

The middle NS5'-brane is located at $x^6=0$ and the $x^6$ 
coordinates for the $NS5_L'$-brane, $NS5_L$-brane, $NS5_R$-brane and
$NS5_R'$-brane are given  by $x^6=-y_2, -y_1, y_1$ and
$x^6=y_2$
respectively. 
The $N_c$ D4-branes 
are suspended between the 
$NS5_L$-brane, whose $x^6$ coordinate is given by $x^6=-y_1$,  and 
$NS5_R$-brane, whose $x^6$ coordinate is given by $x^6=y_1$,  while 
the $N_c'$ D4-branes 
are suspended between the $NS5_L$-brane and the $NS5_L'$-brane
and further they are 
suspended between the $NS5_R'$-brane and the $NS5_R$-brane.
We draw this brane configuration in Figure 5A for the vanishing mass
for the bifundamentals. See also 
the relevant previous work appeared in 
\cite{Ahn07-4}
\footnote{This is equivalent to the reduced brane
configuration in section 4 of \cite{Ahn07-4} with particular rotations for the
NS5-branes if we remove all the D6-branes 
completely.}.
The gauge couplings of $SU(N_c)$ and $ SU(N_c')$
are given by (\ref{couplings}), as before.
See also the relevant works in \cite{LLL1,BHKL,EGKT}.

\begin{figure}[ht]
   \epsfxsize=5.0in 
\centerline{\epsffile{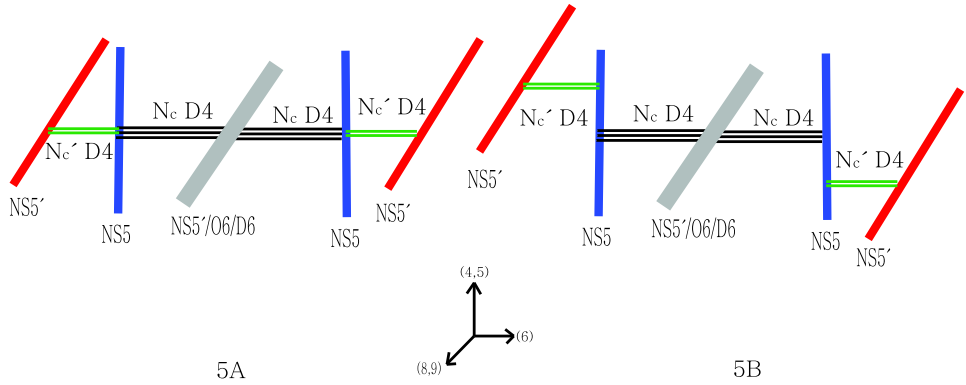}}
   \caption[FIG. \arabic{figure}.]{ 
The  ${\cal N}=1$ supersymmetric 
electric brane configuration for the gauge group $SU(N_c) \times
SU(N_c')$ and the bifundamentals $X$ and $\widetilde{X}$  
as well as antisymmetric, conjugate symmetric flavors and eight D6-branes 
with vanishing(5A) and 
nonvanishing(5B) mass
for the bifundamentals.  }
\end{figure}

There is no electric
superpotential corresponding to the Figure 5A. Now let us deform 
this theory. 
Displacing the two NS5'-branes relative each other in the $v$ 
direction corresponds to turning on a quadratic
superpotential
for the bifundamentals $X$ and $\widetilde{X}$ as (\ref{superelectric1})
where the $\Phi'$
is a meson field
and the mass $m$ is given by (\ref{mass1}).
The $NS5_L'$-brane is moving to the $+v$ direction and the 
$NS5_R'$-brane is moving to $-v$ direction due to the O6-plane for
fixed NS5-branes. In other words,
the $x^5$ coordinate of $NS5_L'$-brane is $+ \Delta x$ while
the $x^5$ coordinate of $NS5_R'$-brane is $- \Delta x$.
We draw this brane configuration in Figure 5B for nonvanishing mass
for the bifundamentals
by moving the $NS5_L'$-brane with 
$N_c'$ color D4-branes 
to the $+v$ direction and their mirrors to $-v$
direction. 


Let us apply the Seiberg dual to the $SU(N_c)$ factor.
Starting from Figure 5B and moving the $NS5_L$-brane to the right all the
way past the $NS5_M'$-brane(and $NS5_R$-brane to the left of $NS5_M'$-brane),
one obtains the Figure 6A.
By introducing $N_c'$ D4-branes and $N_c'$ 
anti-D4-branes  between $NS5_R$-brane and   
$NS5_M'$-brane, 
we are left with 
$(N_c'-\widetilde{N}_c)$ anti-D4-branes between $NS5_R$-brane and   
$NS5_M'$-brane.
The brane configuration for zero mass for the bifundamental
can be obtained from Figure 6A by 
pushing $N_c'$ D4-branes into the origin $v=0$.

\begin{figure}[ht]
   \epsfxsize=5.0in 
\centerline{\epsffile{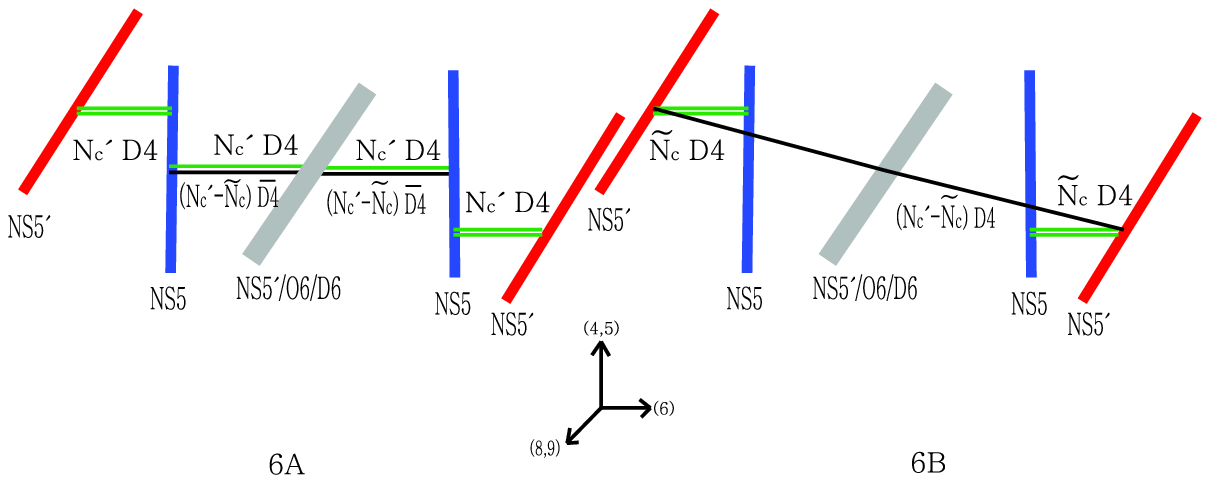}}
   \caption[FIG. \arabic{figure}.]{ 
The  magnetic brane configuration corresponding to Figure 5B with D4-
and $\overline{D4}$-branes(6A) and  with 
a misalignment between D4-branes(6B)  when the NS5'-branes are close to
each other. }
\end{figure}

The gauge group is given by
\bea
SU(\widetilde{N}_c=2N_c'-N_c+4) \times SU(N_c')
\label{dualdual}
\eea
where the number of dual color can be obtained from the linking number 
counting, as done in \cite{Ahn07-4,Ahn07-1}.
The matter contents are the flavor singlet $Y$ in the bifundamental 
representation $({\bf \widetilde{N}_c, \overline{N_c'}})$
and its complex conjugate field $\widetilde{Y}$ in the bifundamental 
representation  $({\bf \overline{\widetilde{N}_c}, N_c'})$,
and  the gauge singlet $\Phi' \equiv X \widetilde{X}$ 
in the representation for 
$({\bf 1, N_c^{'2}-1}) \oplus ({\bf 1, 1})$, under the dual gauge group.
There are also
the antisymmetric flavor $a$, the conjugate 
symmetric flavor $\widetilde{s}$ and eight fundamentals $\hat{q}$ for 
$SU(\widetilde{N}_c)$.

Then the dual magnetic superpotential, by adding the mass term
 for the bifundamental $X$, is given by
\bea
W_{dual} =  \Phi' Y \widetilde{Y}  + m \Phi'
+ \hat{q}
 \widetilde{s} \hat{q} 
\label{super}
\eea
where this can be seen from the 
equation (4.2) of \cite{Ahn07-4} by putting the terms coming from  
the D6-branes in both electric and magnetic theories to zero.

The brane configuration in Figure 6A is stable as long as the
distance $\Delta x$ between the upper $NS5_L'$-brane and 
the middle $NS5_M'$-brane is large. If they are close to 
each other then this brane
configuration is unstable to decay and it becomes 
the brane configuration in Figure
6B.
Since the two NS5'-branes are located at different sides of NS5-brane
in Figure 6B, contrary to the previous cases,
the $x^6$ coordinates for NS5'-branes are positive and negative when 
we take $x^6=0$ for the NS5-brane.   
For the DBI computation, this fact should be taken into account. 
One can regard these brane configurations as particular states in the
magnetic gauge theory with the gauge group (\ref{dualdual}) and
superpotential (\ref{super}).
When the two NS5'-branes which are connected by $\widetilde{N}_c$
D4-branes 
are replaced by two coincident
D6-branes,
the brane configuration of Figure 6B is the same as the one studied in 
\cite{Ahn07-1,FGU}.

The gauge couplings for the two gauge group factors are
given by
\bea
g_{1,mag}^2 = \frac{g_s \ell_s}{y_1}, \qquad 
g_{2,mag}^2 = \frac{g_s \ell_s}{(y_2-y_1)}
\nonu
\eea
and the superpotential 
corresponding to Figures 6A and 6B is given by
\bea
W_{dual} = h \Phi' Y \widetilde{Y} - h \mu^2 \Phi' + \hat{q}
 \widetilde{s} \hat{q} , \qquad
h^2=  g_{2,mag}^2
\nonu
\eea
and the  mass parameter $\mu^2$ is given by (\ref{mass}).
Then $ Y \widetilde{Y}$ is 
a $\widetilde{N}_c \times \widetilde{N}_c$ 
matrix where the second gauge group indices for $Y$ and $\widetilde{Y}$ 
are contracted with those
of $\Phi'$ while $\mu^2$ is a 
$N_c' \times N_c'$ matrix.
Although the field $Y$ itself is a fundamental in the second gauge
group
which is a different feature, compared with the singlet 
representation for the usual quark
coming from D6-branes \cite{Ahn07-4},
the product $Y \widetilde{Y}$ has the same representation with the 
product of dual quarks, $q \widetilde{s} a \widetilde{q}$ in the notation of 
\cite{Ahn07-4}.
Moreover, the second gauge group indices for the field $\Phi'$ play the
role of the flavor indices for the gauge singlet $M' \equiv
Q \widetilde{Q} $ in \cite{Ahn07-4}.

Therefore, the F-term equation, the derivative $W_{dual}$ with respect to the
meson field $\Phi'$ cannot be satisfied if the $N_c'$ exceeds
$\widetilde{N}_c$.
So the supersymmetry is broken.   
The classical moduli space of vacua can be obtained from F-term
equations. 
That is, 
there are five equations from F-term conditions:
$
Y \widetilde{Y} -\mu^2 =0,  \Phi' Y =0,  \widetilde{Y}
\Phi'=0, \hat{q} \widetilde{s} =0$, and $ \hat{q} \hat{q} =0$.
Then the solutions for these
are given by 
\bea
<Y>   & = & 
\left(
\begin{array}{c}
\mu  e^{\phi} {\bf 1}_{\widetilde{N}_c}  \nonu \\
0
\end{array}
\right), \quad
<\widetilde{Y}>   = 
\left(
\begin{array}{cc}
\mu  e^{-\phi} {\bf 1}_{\widetilde{N}_c} & 0 \nonu \\
\end{array}
\right), 
\quad
<\Phi'> =
 \left(
\begin{array}{cc}
0  & 0
 \\
0 & \Phi_0  {\bf 1}_{(N_c'-\widetilde{N}_c)} 
\end{array}
\right),
\nonu \\
<\hat{q}> & = & 0,   \quad 
<\widetilde{s}> = 0.
\nonu
\eea
One can expand around the solutions. Although there
exists an extra last term  in (\ref{super}), this does not contribute to the
one loop result.
At one loop, the effective potential $V_{eff}^{(1)}$ for $\Phi_0$
leads to the positive value for $m_{\Phi_0}^2$ implying that these
vacua are stable.
The gauge theory analysis where the theory will be strongly coupled in
the IR region $N_c' > 2\widetilde{N}_c-4$ is only valid in the regime where 
$\Delta x$ is smaller than $\exp(-\frac{C}{g_s})$ with some positive
constant $C$ as before. 

\section{Conclusions and outlook}

The meta-stable brane configurations we have found are summarized by
Figures 2, 4, and 6.
If we replace the upper and lower NS5'-branes in Figures 2B, 4B and 6B
with the coincident D6-branes, 
those brane configurations become nonsupersymmetic
minimal energy brane configurations in \cite{FGU,Ahn06-1}, 
in \cite{FGU,Ahn07}, and in \cite{FGU,Ahn07-1} respectively.

It would be interesting to 
construct the meta-stable brane configuration where
there exist four NS5-branes by adding one extra outer NS5-brane  to
the brane configuration found in \cite{GK} or to the brane configuration 
of Figure 1 in this paper or to
construct the meta-stable brane configuration where
there exist six NS5-branes by adding two extra outer NS5-branes  to
the brane configuration found in \cite{Ahn07-3}.
Or one can add  two extra outer NS5-branes  to
the brane configuration found in \cite{Ahn07-4}.
These gauge theories will be  triple product gauge group theories.  

Some different directions concerning on the meta-stable vacua
in different contexts are present in
recent works \cite{MPS}-\cite{ABFK} where some of them use anti D-branes 
and some of them are described in the type IIB theory.
It would be very interesting to find out
how the meta-stable brane configurations from type IIA string theory 
and those brane configuration from type IIB theory are related to each other.

\vspace{.7cm}

\centerline{\bf Acknowledgments}

I would like to thank 
D. Kutasov 
for discussions and Harvard High Energy Theory Group for hospitality
where part of this work was undertaken. 
This work was supported by grant No.
R01-2006-000-10965-0 from the Basic Research Program of the Korea
Science \& Engineering Foundation.

\end{document}